\documentclass[useAMS,usenatbib]{mn2e}
\usepackage{epsf}
\usepackage{amsmath,amssymb}
\usepackage{graphicx}

\title[Asymmetric Explosions of Thermonuclear Supernovae]{Asymmetric Explosions of Thermonuclear Supernovae}
\author[C. R. Ghezzi]{C. R. Ghezzi$^{1}$ $^{2}$ \thanks{E-mail:
ghezzi@ime.unicamp.br},
E. M. de Gouveia Dal Pino$^{2}$ and 
J. E. Horvath$^{2}$
\\
$^{1}$Departamento de Matem\'atica, Instituto de Matem\'atica, Estat\'{\i}stica e \\
Computa\c c\~ao Cient\'{\i}fica, Universidade Estadual de Campinas, 13081-970, Campinas, SP, 
Brasil.\\
$^{2}$Instituto de Astronomia, Geof\'{\i}sica e Ci\^encias Atmosf\'ericas,
Universidade de S\~ao Paulo,\\
R. do Mat\~ao, 1226, Cd. Universit\'aria,
05508-090,  \\ S\~ao Paulo, SP,  Brasil}

\date{Accepted ..... 
      Received ..... ; 
      in original form .....}

\begin{document}

\date{}

\pagerange{\pageref{firstpage}--\pageref{lastpage}} \pubyear{2002}

\maketitle

\label{firstpage}

\begin{abstract}
A type Ia supernova explosion starts in a white dwarf as a laminar
deflagration at the center of the star and soon several hydrodynamic
instabilities (in particular, the Rayleigh-Taylor (R-T) instability)
begin to act.
In  previous work (Ghezzi, de Gouveia Dal Pino, \& Horvath
 2001), we addressed the propagation of an initially laminar
thermonuclear flame in presence of a magnetic field assumed to be
dipolar. We were able to show that, within the framework of a
fractal model for the flame velocity, the front is affected by the
field through the quenching of the R-T instability growth in the
direction perpendicular to the field lines. As a consequence, an
$asymmetry$ develops between the magnetic polar and the equatorial
axis  that gives a prolate shape to the burning front. We have
here computed numerically the total integrated asymmetry as the
flame front propagates outward through the expanding shells of
decreasing density of the magnetized white dwarf  progenitor, for
several chemical compositions, and found that a total asymmetry of
about 50\% is produced between the polar and equatorial directions
for  progenitors with a surface magnetic field
B $ \sim 5 \times 10^{7}$ G, and a composition $^{12}{\rm C} = 0.2$ and
$^{16}{\rm O} = 0.8$  (in this case, the R-T instability saturates at scales
$\sim$ 20 times  the width of the flame front). This asymmetry is
in good agreement with the inferred asymmetries from
spectropolarimetric observations of very young supernova remnants,
which have recently revealed intrinsic linear polarization
interpreted as evidence of an asymmetric explosion in several
objects, such as SN1999by, SN1996X, and SN1997dt. Larger magnetic
field strengths will produce even larger asymmetries. We have also
found that for $lighter$ progenitors (i.e., progenitors with smaller concentrations of
$^{16}$O and larger concentrations of $^{12}$C) the  total asymmetry is
larger.
\end{abstract}


\section{Introduction}

According to the current modelling, the explosion of a type Ia
supernova starts as a thermonuclear deflagration of a
Chandrasekhar mass white dwarf of carbon-oxygen (C+O)  or
oxygen-neon-magnesium (O+Ne+Mg) compositions. Due to the action of
several hydrodynamic instabilities, in particular, the
Rayleigh-Taylor (R-T), the initially laminar propagation does not
survive as such and the combustion front rapidly develops a
complex topology. A cellular stationary combustion (followed by a
turbulent combustion regime) is rapidly achieved by the flame and
maintained up to the end of the so-called flamelet regime when a
transition to detonation may occur. As in the case of chemical
laboratory flames (Gostintsev et al 1988), thermonuclear flames
probably develop a self-similar behavior due to the action of the
hydrodynamic instabilities. Under this hypothesis of a
self-similar deflagration regime, the effective flame speed is
well described by a fractal scaling law, as previously proposed by
several authors (Woosley 1990, Timmes \& Woosley
 1992, Niemeyer \& Hillebrandt 1995,  Niemeyer \& Woosley 1997,
 see also Ghezzi, de Gouveia Dal Pino \& Horvath 2001), namely
\begin{equation}
\label{fracvel}
v_{frac}=v_{lam} \biggl(\frac{L_{max}}{l_{min}}\biggr)^{(d-2)}\,,
\end{equation}
where $v_{lam}$ is the laminar velocity  (as obtained, for
example, in Timmes \& Woosley 1992); $L_{max}$ and $l_{min}$ are
the maximum and minimum Rayleigh-Taylor wavelength perturbations,
respectively (see Chandrasekhar 1961); and $d$ is the fractal
dimension, which may assume a range of values, $2 \le d < 3$
(Woosley 1990; see also Ghezzi, de Gouveia Dal Pino \& Horvath
2001).

In a previous work (Ghezzi, de Gouveia Dal Pino \& Horvath 2001,
hereafter, Paper I), we addressed the propagation of an initially
laminar thermonuclear flame in presence of a magnetic field
assumed to be of dipolar geometry. The main result of that work
was that within the framework of the fractal models for the flame
velocity in the wrinkled and turbulent regimes, the front is
affected by the field that inhibits the growth of the R-T
instability in the direction perpendicular to the field lines. As
a consequence,  an $asymmetry$ is established between the magnetic
polar and the equatorial axis  that gives a prolate shape to the
burning front (see Figure 1). In that work, an estimate of the
intrinsic asymmetry as a function of the field and the core
density was presented and discussed in the context of type Ia
supernova explosions. Nevertheless, several aspects of this
problem needed to be clarified as, for instance, a computation of
the total integrated asymmetry, as the burning front propagates
through the outer, density-decreasing shells of the magnetized
expanding progenitor star, and also its meaning for the
observation of SNIa remnants. We here present the results of
one-dimensional numerical calculations of the integrated asymmetry
through white dwarf progenitors  with different initial chemical
compositions, taking into account the expansion of the star (which
is assumed to be homologous).

\section{Instantaneous and Corrected Asymmetry}

According to the model developed in Paper I,
in the presence of a dipolar magnetic field,
the ratio of the propagation velocity of the burning front in the polar
and equatorial directions of the star is given by

\begin{equation} \label{asiminst}
A(\rho)=\frac{v_{pol}(\rho)}{v_{eq}(\rho)}=\biggl(
1+\frac{B^2/8\pi}{\rho\,v_{lam}^2/2} \biggr)^{d-2}\,,
\vspace{0.3cm}
\end{equation}

\noindent
where $v_{pol}$ and $v_{eq}$ are the velocities at the
polar and equator directions, respectively. This equation shows
that the asymmetry of the velocity field is controlled by the
quotient of  the magnetic pressure ($ B^{2}/8\pi$) and the ram
pressure of the gas ($\propto \rho\,v_{lam}^{2}$). We shall label
$A(\rho)$ as the {\it instantaneous asymmetry} as a function of
the density $\rho$ for an initially spherical flame. In paper I,
we have computed this quantity for different magnetic field
intensities.

It is clear, however, that $A(\rho)$ will be increasingly
different from the quotient above as time elapses and the flame
propagates outward through the star, since the flame in both
directions will be encountering decreasing densities ahead.
Therefore, a full integration of the asymmetry is needed to
account for this fact.

Let us first assume that the star does not expand to derive a
zero-th order correction. Since the polar flame front is faster
than the equatorial front, the former will encounter lower
densities first. If label the difference of densities in each
direction as $\Delta \rho=(\rho_{eq}-\rho_{pol})>0$, the polar
front faces a density $\rho_{pol}=\rho-\Delta \rho $ when the
equatorial front is at $\rho_{eq}=\rho$. To this order, a {\it
corrected asymmetry} (which takes into account the deformation of
the flame) may be defined by

\begin{equation}
\mbox{\it \~A}(\rho)=\frac{v_{pol}(\rho-\Delta
\rho)}{v_{eq}(\rho)}\,.
\end{equation}

Since the fractal propagation velocity is a continuous and
differentiable function of the density, we can expand $v_{pol}$ in
a Taylor series around $\rho$ and keep the zeroth and first order
terms provided that  $\Delta \rho \ll \rho$, to yield

\begin{equation} 
\mbox{\it \~A}(\rho)=\frac{v_{pol}(\rho)-
\frac{\partial\,v_{pol}\,(\rho)}
{\partial\,\rho}\,\, \Delta \rho}{v_{eq}(\rho)} \,,
\end{equation}

\noindent which may be rearranged to display the instantaneous
asymmetry and a correction term as

\begin{equation}
\label{asycorr}
\mbox{\it \~A}(\rho)=A(\rho)-\biggl[
\frac{1}{v_{eq}}\, \frac{\partial\,v_{pol}\,(\rho)}{\partial\,\rho}
\biggr]
\,\,\Delta \rho
\equiv A(\rho)+A_c(\rho)\,,
\vspace{0.3cm}
\end{equation}
where
\begin{equation} \label{correc}
A_c(\rho)=-\biggl[
\frac{1}{v_{eq}}\, \frac{\partial\,v_{pol}\,(\rho)}{\partial\,\rho}
\biggr]
\,\,\Delta \rho\,\,,
\end{equation}
is the correction term for the asymmetry (see below).

Given that $v_{pol} > v_{eq}$ and both velocities increase with
decreasing density \footnote{The fractal velocity, $v_{frac}$,
increases with the density, in Eq. (1), because the minimum scale
$l_{min}$ decreases faster with density than the laminar velocity,
$v_{lam}$ (see Paper I).}, the difference between $v_{pol}$ and
$v_{eq}$ is necessarily larger than the difference that they would
have by imposing the same density ahead for both flames. In other
words, the {\it corrected asymmetry} $\mbox{\it \~A}(\rho)$ must be larger
than the instantaneous asymmetry $A(\rho)$. Moreover, the value of
$A_{c}(\rho)$ depends on the value of $\Delta\rho$, which in turn,
depends on the value of the corrected asymmetry $\mbox{\it \~A}(\rho)$
defined in Eq.(5). We may then find a solution to the non-linear
Eq. (5) by discretizing and solving it iteratively \footnote{It
can be checked that the Lipschitz condition is satisfied by the
function $\mbox{\it \~A}(\Delta \rho)$. }

\begin{equation} \label{iterar}
\mbox{\it \~A}_{n}(\rho)=A_{n-1}(\rho)-\biggl(\frac{v'_{pol}}
{v_{eq}} \biggr)_{n-1} \,\,\Delta \rho_{n-1}\,,
\end{equation}
with  $\mbox{\it \~A}_o=A(\rho)$, and  $\Delta \rho_o=0$.
Whenever $\Delta \rho \sim \rho$, higher order terms become important,
although $\rho \gg \Delta \rho$ is automatically satisfied.
 Eq.(7) converges provided that the Cauchy  condition
\begin{equation} \label{converg}
| \mbox{\it \~A}_{n}(\rho) - \mbox{\it \~A}_{n-1}(\rho)|\,< < \epsilon
\end{equation}

\noindent
is satisfied for an arbitrarily small $\epsilon$.
We shall see below how to integrate the corrected asymmetry through the
star so as to estimate the accumulated
effect along the whole propagation of the non-spherical flame.

\subsection{Correction Term of the Asymmetry}

The correction term for the asymmetry  (Eq. 6), contains a
derivative of the polar speed with respect to the density, and
therefore, we need to know the dependence of the laminar speed
with the density and composition of the fuel in order to calculate
the correction term. In this subsection we calculate the
correction term for C$+$O progenitors. Using the interpolation
formulae given by Woosley (1986); Timmes \& Woosley (1992); and
Arnett (1996), we find

\begin{equation}
\label{vlamrho}
v_{lam}=92 \times 10^5\,\,\biggl( \frac{\rho}{2\times10^9}\biggr)
^{0.805}\,\biggl[ \frac{X(^{12}{\rm
C})}{0.5}\biggr]^{0.889}
\end{equation}
$$\hspace{6cm}{\rm cm\,s}^{-1},$$

which is approximately\footnote{We should note that equation (9) is only approximate and it is used for
simplicity reasons, more realistic results could probably be obtained directly interpolating the velocities 
given in table ($3$) of Timmes \& Woosley 1992, and will be adressed in future calculations.}
 valid in the density range \mbox{$0.01 \le
\rho_{9} \le 10$}. In the  particular case in which $d=2.5$
\footnote{This is actually the maximum fractal dimension estimated
in Paper I.}, the  effective polar speed $v_{pol}$ (see Eq. 1)
does not depend on the laminar speed $v_{lam}$, so that  we could
say that $v_{pol}$ is to some extent ``independent of the
microphysics"\footnote{ In fact, recent numerical simulations of
supernovae have shown that the hydrodynamics of the explosion is
independent of the microphysics of the flame (Gamezo et al. 2002).
In paper I,  we found that the minimum relevant hydrodynamic scale
for the flame was $l_{min} \sim 10^5\,$cm (for
$\rho \sim 10^9\,{\rm g\,cm}^{-3}$), and the flame evolution was "independent
of the microphysics"  since $l_{min} \gg \delta_f$ almost all the
way, where $\delta_f$ is the flame width. We here stop the
numerical integration when $l_{min} \sim \delta_f $ or $l_{min}
\sim 20 \times \delta_f$ (see below). In the particular case that
$d=2.5$, using the fractal scaling given by eq. (1) we recover 
the independence with microphysics which was
obtained through dimensional arguments by Khokhlov (1995), and is
also consistent with laboratory  results of Taylor (see Kull
1991). In this case, the flame does not depend directly on
quantities, such as, the laminar velocity or the thermometric
conductivity.}

\begin{equation}
v_{pol}(\rho_{u})=0.282\,\rho_{u}^{0.8}\,\biggl(\frac{g\, L_{max}\,
\delta
\rho}{\rho_{u}^{2.6}} \biggr)^{0.5}\,\,\,\,\,\,\,{\rm cm\,s}^{-1}\,.
\end{equation}
Here, $\delta \rho=\rho_{u}-\rho_{b}$ is the density jump at the flame,
i.e., the difference between burned  ($\rho_{b}$)  and unburned ($\rho_{u}$)
material densities.
We note that this is, in general,  different from the
 the density difference between the polar and equatorial fronts
$\Delta \rho=\rho_{eq}-\rho_{pol}$ (see Eqs. 3 - 7).
Thus

\begin{equation}
\frac{\partial v_{pol}}{\partial
\rho_u}=\frac{0.225\,\,\biggl(\frac{g\,L_{max}\,\delta
\rho}{\rho_{u}^{2.6}}
\biggr)^{0.5}}{\rho_{u}^{0.2}} -
\end{equation}
$$
\frac{0.141\,\,\rho_{u}^{2.1}\,\,\biggl(\frac{2.6\,g\,L_{max}\,\delta
\rho}{\rho_{u}^{3.6}}+\frac{g\,L_{max}}{\rho_{u}^{2.6}}
\biggr)}{(g\,L_{max}\,\delta \rho)^{0.5}}\,\,.
$$

Using   Eqs. (1) and (9), it is possible to obtain the polar
fractal speed, $v_{pol}$, for any C$+$O composition with an
arbitrary fractal dimension $d$

\begin{equation}
v_{pol}=14.1683\,\,(0.208752)^d\,\,X(^{12}{\rm
C})^{0.889}\,\,\times
\end{equation}
$$\rho_{u}^{0.8}\,\,\biggl[\frac{g\,L_{max}\,\delta
\rho}{X(^{12}{\rm C})^{1.778}\,\,\rho_{u}^{2.6}}\biggr]^{d-2}\,.$$

And, in general, the correction term is given by

\begin{equation}
\frac{1}{v_{eq}} \frac{\partial v_{pol}}{\partial \rho_u} \Delta \rho=
\frac{1}{v_{eq}\,g^2\,L_{max}^2\,\delta
\rho^3}\biggl[22.6693\,\times
\end{equation}
$$(0.208752)^d\,X(^{12}{\rm C})^{4.445}\,\biggl(\frac{g\,L_{max}\,\delta \rho}{X(^{12}{\rm
C})^{1.778}\,\rho_{b}^{2.6}}\biggr)^d \,\times
$$
$$
\biggl(-1.625\,(
d-2.30769 )\, \rho_{u}\,\rho_{b}^5 +( d-2.5 )\,\rho_{b}^6 \biggr)
\biggr]\, \Delta \rho\,\,; $$

\noindent where, the value of $\Delta \rho$ must be calculated
numerically (see the next section).

\section{Integrated Asymmetry}

In order to compute the total asymmetry for a given set of initial
conditions for the progenitor, we must integrate the following
equations along both the equatorial and polar directions,
respectively

\begin{equation}
\label{req}
\frac{d r_{eq}}{dt}=v_{eq}(\rho,t)
\end{equation}

\begin{equation}
\label{rpol}
\frac{d r_{pol}}{dt}=v_{pol}(\rho,t)
\end{equation}

\noindent
where $v_{pol}(\rho) = \mbox{\it \~A}(\rho) \, v_{eq}(\rho)$,
and $v_{eq}$ is given by:

\begin{equation}
\label{fraceq}
v_{eq}=v_{lam} \biggl(\frac{L_{max}}{l_{eq}}\biggr)^{(d-2)}\,.
\end{equation}

\noindent
where $l_{eq}=l_{min}$ at the magnetic equator of the star (Paper I):
\begin{equation}
\label{lambdaeq}
\l_{eq}= \frac{8 \pi \biggl(\frac{B^2}{8 \pi}+
\frac{1}{2} \rho \, v_{lam}^{2} \biggr)}{g \,\delta \rho}\,\,,
\end{equation}

The equations above  can be integrated simultaneously in one dimension
outward through the star along with Eqs. (7) and (8) to give the
integrated asymmetry $A_{tot}(\rho)= r_{pol}(\rho)/r_{eq}(\rho)$,
up to a shell of   density $\rho$.
The integration of Eq.  (14) is straightforward, while Eq. (15)
is  non-linear since the value of $\mbox{\it \~A}$ depends on the actual position of the polar front
$r_{pol}$.  It
is solved using the same iterative algorithm that gives  the {\it corrected asymmetry}
described in the previous section.

Until now, we have neglected the fact that the flame is actually
propagating in an expanding star and have thus partially decoupled
the equations of the burning front from the hydrodynamical motions
of the full star. Since actual deflagrations are subsonic, the
star will  pre-expand ahead of the flame, and this will, in turn,
affect the burning itself (see Figure 1).

To a very good approximation, we can assume
an {\it homologous} expansion of the white dwarf and this
 allows a simple description of the expansion as a function of time,
in which the radius of the $n$-th zone at a given time $t$ evolves
according to

\begin{equation} \label{exphom}
r_{n}(t)=a(t)\,r^o_{n}\,,
\end{equation}

\noindent
where $r^o_{n}$ gives the distance to the center in the initial
condition (i.e., in hydrostatic equilibrium), and $a(t)$ is the homology factor satisfying
$a(0) = 1$.
The density will then evolve according to

\begin{equation}
\rho_{n}(t)=\rho^{o}_{n}/a^3(t)\,,
\end{equation}
\noindent where $\rho_n^{o}$ is the density at the n-th zone in
hydrostatic equilibrium, so that in the comoving frame of each
zone the density will decrease with time. \footnote{We note that
some deviation of a perfect homologous expansion should be
expected mainly due to the fact that, as the flame is accelerated
downstream under the action of hydrodynamical instabilities, shock
and pressure waves may travel ahead of the flame and disturb
locally the stellar structure. Nonetheless, since the deflagration
evolves under approximate isobaric conditions, the actual
expansion is expected to be nearly close to the homology  assumed
here.}

It is easy to see that the stellar expansion should produce a
slower flame front than in the stationary case. This is due to the
fact that when the deflagration reaches the radius $r^{n}$,the
density there will be lower than it is without expansion.
Therefore, the propagation velocity will be modified accordingly.
The distance between zones is now $dr_{eq}' = a(t)\,dr_{eq}$, and
$dr_{pol}'= a(t)\,dr_{pol}$, for each of the propagation
directions. The $n$-th zones evolve according to
$${r_{eq}}_n'={r_{eq}}_{n-1}'+{dr_{eq}}_{n-1}'\,,$$
and
$${r_{pol}}_n'={r_{pol}}_{n-1}'+{dr_{pol}}_{n-1}'\,,$$
respectively.

\begin{figure}
\begin{center}
\includegraphics[width=80mm]{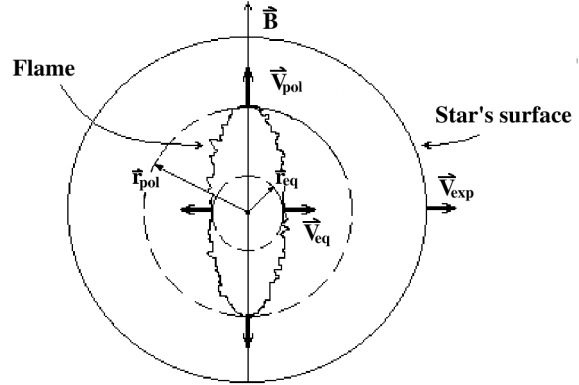}
\caption{
Schematic representation of the propagation of the front inside a
white dwarf. Here $\vec{{\rm B}}$ indicates the direction of the dipolar
magnetic field,  $\vec{{\rm V}}_{exp}$ is the star's expansion velocity,
$\vec{{\rm V}}_{pol}$ and  $\vec{{\rm V}}_{eq}$ are the fractal polar and
equatorial velocities, and ${\rm r}_{pol}$ and ${\rm r}_{eq}$ are the polar
and equatorial radii, respectively, as  described in the text. We
note that the combustion is isobaric to a very good approximation,
and the density jump in the flame is $\sim 10^{-1}$ times the
density value.}
\end{center}
\end{figure}

We have  computed the total asymmetry  by taking into account the
stellar expansion effect upon the density, for different initial
chemical compositions  of the  white dwarf progenitor. We have
integrated Eqs. (14) and (15) together with Eqs. (18) and (19),
using a subroutine to interpolate step by step the physical
properties of the deflagration front, i.e., the laminar velocity
and the density jump, and the value of the magnetic field. We have
employed an analytical function for $a(t)$, namely
$a(t) = 0.25\, t + 1$ obtained from a numerical fit to the  homologous expansion
solutions of Goldreich \& Weber (1980). The magnetic field
strength has been assumed to decrease approximately linearly with
the radial coordinate (see Wendell et al. 1987, and Paper I), and the
initial density and gravity of the  progenitor are depicted in
Figures 2 and 3, respectively, as  functions of the stellar
radius. To compute the variation of the magnetic field in the
comoving frame of the flame,  we have assumed magnetic flux
conservation. The computed  progenitor models were resolved in
$10^{3}$ zones, giving a physical resolution $\sim 1.5 \,$km for
a star with a radius R$ = 10^{5} $km. The integration was stopped
when the polar combustion front reached a shell with density $\rho
\sim 10^7$ g cm$^{-3}$ for which  the transition to detonation is
believed to occur (Khokhlov 1995). This is a first step towards a
more complete calculation, but nevertheless contains all the
essential ingredients necessary to address the total asymmetry
$A_{tot}'$, namely

\begin{equation}
A_{tot}'=r_{pol}'/r_{eq}'\,\,,
\end{equation}
where, generally speaking, $r_{eq}'>r_{eq}$ and $r_{pol}'>r_{pol}$.

\begin{figure}
\vspace{0.5cm}
\begin{center}
\includegraphics[width=84mm]{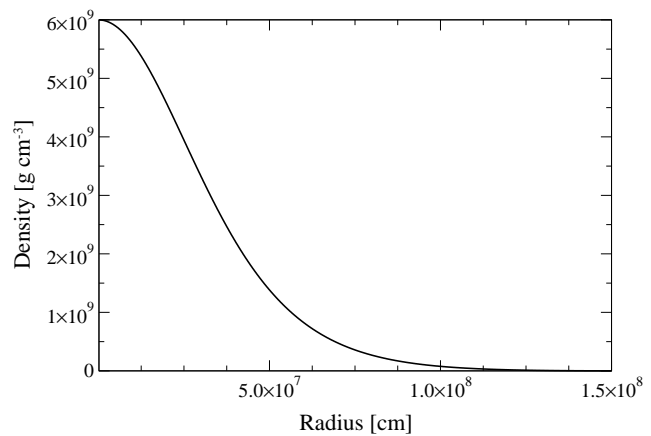}
\caption{Initial density of the white dwarf progenitor as a function of
the stellar radius.}
\end{center}
\end{figure}

\begin{figure}
\vspace{1cm}
\begin{center}
\includegraphics[width=84mm]{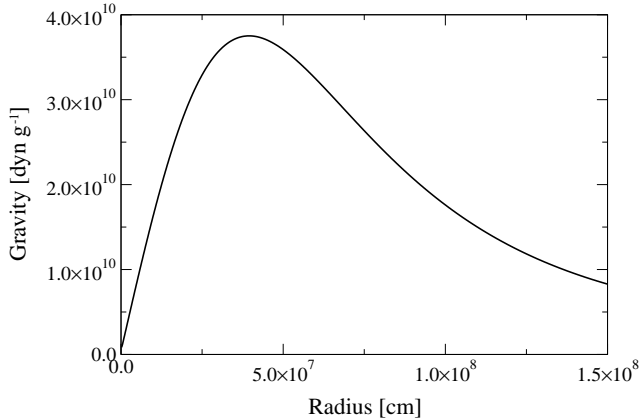}
\caption{Initial absolute value of the gravity of the white dwarf
progenitor as a function of the stellar radius calculated for a
progenitor with an electron fraction ${\rm Y_e}=0.5$.}
\end{center}
\end{figure}

\section{Results}

The results of the calculations are shown in Figures 4 to 8.
The instantaneous asymmetry near the stellar surface
(for a density $\rho = 5 \times 10^7\,\,{\rm g\,cm}^{-3}$)
is shown in Figure 4 for several progenitors with C-O
compositions as a function of the surface magnetic field.

Integrated asymmetries are depicted in Figure 5 for
three different compositions (see captions) and a  maximum $surface$
magnetic field  B$ = 10^{9}$ G. It is apparent
from Figs. 4 and 5 that heavier
progenitors develop larger instantaneous asymmetries,
but the lighter ones achieve the largest integrated asymmetries. Hence, it is important to understand the
reasons behind such behavior for future applications.

The key aspect for this behavior may be traced back to saturation
effects of the flame as follows. The value of $l_{min}$ (see Eq.
17) indicates that it must decrease as the density decrease, but
it clearly cannot be arbitrarily small. In fact,  there is a
minimum value associated with the width of the flame $\delta_{f}$,
which we will call the saturation  scale, $l_{sat}$.

\begin{figure}
\vspace{1cm}
\begin{center}
\includegraphics[width=84mm]{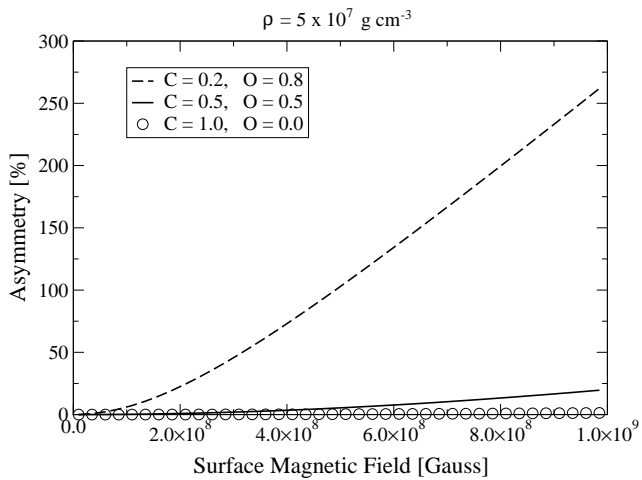}
\caption{Instantaneous asymmetry for progenitors with different
initial compositions  [
 (a)
$X(^{12}{\rm C})=0.2$, $X(^{16}{\rm O})=0.8$; (b) $X(^{12}{\rm C})=0.5$,
$X(^{16}{\rm O})=0.5$;
and (c) $X(^{12}{\rm C})=1.0$, $X(^{16}{\rm O})=0.0$] as a function of the magnetic field at a
density $\rho = 5 \times 10^7\,\,{\rm g\,cm}^{-3}$.}
\end{center}
\end{figure}

\begin{figure}
\vspace{1cm}
\begin{center}
\includegraphics[width=84mm]{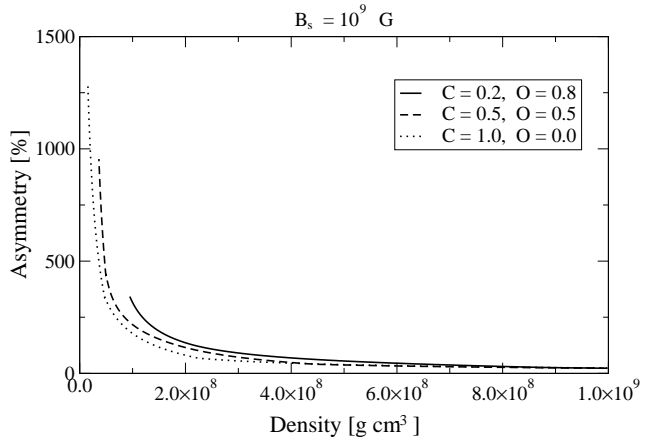}
\caption{Integrated asymmetries for the progenitors (a)
$X(^{12}{\rm C})=0.2$, $X(^{16}{\rm O})=0.8$; (b) $X(^{12}{\rm C})=0.5$,
$X(^{16}{\rm O})=0.5$; (c) $X(^{12}{\rm C})=1.0$, $X(^{16}{\rm O})=0.0$. A surface
field B$ = 10^9$ G is assumed for all cases (which corresponds
to a core B$ \simeq 10^{10}\,$ G), and a saturation scale of
$20\,\delta_{f}$. Taking lower values for the saturation scale
higher integrated asymmetries are obtained. A general trend of the
asymmetry effect is seen here, i.e., lighter progenitors show
higher integrated asymmetry (while heavier progenitors have higher
instantaneous asymmetry, for a given density and magnetic field,
as shown in Fig. 4). This graph can be also interpreted as a
display of the evolution history of the asymmetry. In the
beginning,  when the flame is still propagating in a high density
medium, the asymmetry is very low. Later on, near the end of the
deflagration regime, the asymmetry largely increases until the
front encounters very low densities and the flame vanishes.}
\end{center}
\end{figure}

The asymmetry between the polar and equatorial directions begins
to decrease when the instabilities in the polar front reach the
saturation limit $l_{sat}$.  The polar front will then stop
accelerating, while the equatorial front will be still
accelerating. This will give place  to a symmetrization phase that
will last until the flame is quenched, or a transition to
detonation occurs. We cannot predict exactly this symmetrization
effect here, but if a transition to detonation occurs at densities
$\sim 10^7\,{\rm g\,cm}^{-3}$, then the errors in our present
calculation must be small. On the other hand, if no transition to
detonation  takes place, then the uncertainties in our estimates
will depend upon the density at which the flame quenches and will
be as large as  small  the  density is. Due to the same reason, we
predict smaller symmetrization effects on lighter progenitors,
i.e., progenitors with $X(^{12}{\rm C})=0.5$ and $X(^{16}{\rm O})=0.5$ will
reach the saturation limit later on at lower densities, and will,
therefore, develop larger integrated asymmetries (see Fig.  5).

It is known that if $l_{sat}= l_{min}\sim \delta_{f}$
thermodiffusive effects must be taken into account  and, in fact,
in most of the laboratory flames $l_{sat} \ge 20\,\delta_{f}$ is
observed (see Bychkov et al. 1999). Since no definitive answer to
this question exists,
we have calculated the integrated asymmetry
for different values of the
saturation scale between these two extremes
(i.e., from $l_{sat} = \delta_{f}$ to
$20\,\delta_{f}$) in Figure 6, for the progenitor (a) of Figure 5.
In Figure 7, we have fixed the saturation scale at
$l_{sat} = 20\,\delta_{f}$, in order to
compare
the integrated and the
instanteneous asymmetry variations with the surface magnetic
field for  the same
progenitor.
Analysing these figures, we find that
the integrated asymmetry is enormous if $l_{sat} \sim \delta_{f}$,
and can be larger than $1000 \%$ if the magnetic field is
$\gtrsim 10^9$ G. When
$l_{sat} \sim 20 \,\delta_{f}$ is imposed, the asymmetries are smaller but
can be still very
large, for instance, the progenitor  shows a maximum
asymmetry of the order of $350 \%$
for $B = 10^9$ G (Fig. 7). When the field is reduced to
$5 \times 10^7$ G, the asymmetry drops. In this
case, it presents a total asymmetry of $\sim 50 \%$,
for $l_{sat} \sim 20\, \delta_{f}$ (Figure 8), which is in the ballpark  of
what polarimetric observations suggest ($\sim 20\,\%$ or less, see
the next section; Leonard et al. 2000, Wang et al. 1997, Howell et
al. 2001).

\begin{figure}
\vspace{0.5cm}
\begin{center}
\includegraphics[width=80mm]{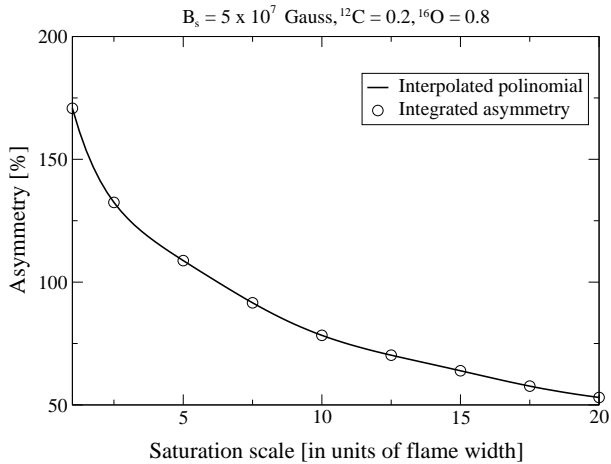}
\caption{
Integrated asymmetry vs saturation scale
$l_{sat}$ in units of the flame width $\delta_f$,
for the  progenitor (a) of Figure 5 with  chemical
composition
$X(^{12}{\rm C})=0.2$,
$X(^{16}{\rm O})=0.8$. A surface field B$ =5 \times 10^7$ G
 is assumed.}
\end{center}
\end{figure}

\begin{figure}
\begin{center}
\includegraphics[width=80mm]{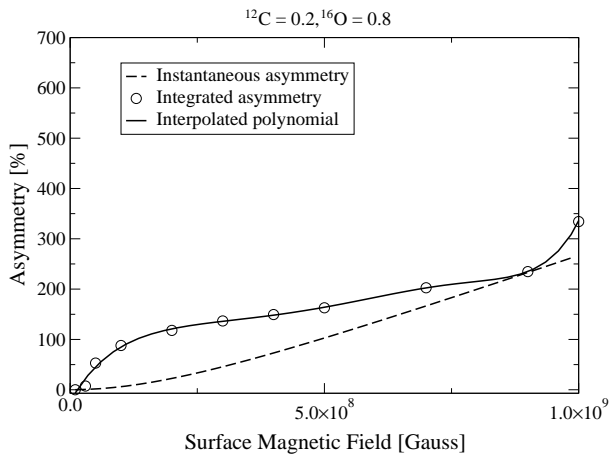}
\caption{
Circles - Integrated asymmetry vs surface magnetic field strength
for the progenitor
with chemical composition $X(^{12}{\rm C})=0.2$,
$X(^{16}{\rm O})=0.8$, and a saturation scale
$l_{sat} = 20\, \delta_f$;
Full line - seven-degree  polinomial curve that fits the numerical results;
Dashed line - instantaneous asymmetry for the same progenitor at a (surface) density
$\rho \sim 10^7\,\,{\rm g\,cm}^{-3}$.}
\end{center}
\end{figure}

\begin{figure}
\vspace{0.5cm}
\begin{center}
\includegraphics[width=80mm]{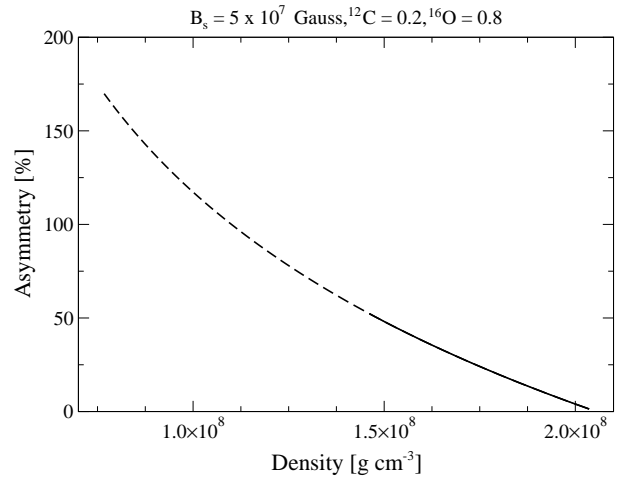}
\caption{Integrated asymmetries for progenitor (a)
$X(^{12}{\rm C})=0.2$,
$X(^{16}{\rm O})=0.8$, with a  surface field B $ = 5 \times 10^7$ G, and
 saturation scales
 $l_{sat} = \delta_f$ (dashed line), and $20 \,\delta_f$ (full line).
This represents the set of parameters that better fits
observational data from SN Ia spectropolarimetric studies and
experimental knowledge from laboratory flames. We have stopped the
calculations when the flame vanishes. At this time, expansion can
start symmetrizing  the flame, although more efficient
symmetrization must be expected  during the coasting phase of the
remnant due to enhanced turbulent diffusion.}
\end{center}
\end{figure}

\section{Discussion and conclusions}

It has been generally assumed  that supernovae explosions are
spherical. However,  recent polarimetric studies of several
supernovae have revealed  intrinsic linear polarization that seems
to  be evidence of  asymmetric  explosion (see Branch et al. 2001;
Leonard et al. 2000; Howell et al. 2001). In the present work, we
have investigated the effects of the magnetic field of the white
dwarf (WD) progenitor on the propagation of the burning front of
thermonuclear supernovae, and found that an asymmetry develops
that could, in principle, help to explain observed asymmetries in
these systems.

The magnetic field strengths inferred for isolated WDs are between
$3 \times 10^4$ G and $10^9$ G, while for WDs in AM Her binaries,
they  are in the range $10^7$ G to $2 \times 10^8$ G
(Wickmasinghe \& Ferrario 2000) and could be further amplified by
compression  during the accretion phase (Cumming 2002). Magnetic
fields had not been taken into account in previous studies of
thermonuclear supernovae because their pressure is found to be
much smaller than the gas pressure. Nonetheless, our studies have
revealed that the magnetic fields act upon the hydrodynamical
instabilities that develop in the flame front (where the magnetic
pressure is larger than the ram pressure; see Eq. 2) and quench
their growth in the direction perpendicular to the field lines. As
a consequence,  an $asymmetry$ develops between the magnetic polar
and equatorial axis  that gives a prolate shape to the burning
front.

We should emphasize that  our  analysis  of the
Rayleigh-Taylor mode interactions at the flame front
has been based on the
model described by Zeldovich
et al. (1966, 1980)\footnote{The Zeldovich et al.'s  model for the
mode-mode interactions of burned cells
explains the saturation of the exponential growth of the hydrodynamic instabilities
 and gives the lower cut-off assumed for the instability  which depends
mainly on the saturation scale as explained above (see  Eq. 17 above and also Eq. 4 of Paper I).},
and its results are in qualitative agreement with  2-D and 3-D numerical
studies of the Rayleigh-Taylor (R-T) instability development in fully turbulent magnetized
layers without combustion (see, Jun, Norman \& Stone 1995, and references therein), and with
numerical simulations of thermonuclear flames in 2D and 3D (see Khokhlov 1995).
Jun et al.'s results, actually
reveal a tendency, in the nonlinear regime,
for an increase of the  growth rate of
the Rayleigh-Taylor instabilities in the direction parallel to
the magnetic field lines (the polar direction),
which would amplify the asymmetry effect here investigated, and
an inhibition  of the growth rate in the direction perpendicular
to the magnetic field lines (equatorial direction)
like in the present investigation.

The calculations performed in the former sections show that the
integrated  asymmetry in the velocity fields may be very large if
the surface magnetic fields are near a maximum value
($\sim 10^9$ G). Observed asymmetries are much less extreme. At least three
alternatives to solve this mismatch come to mind. The first is
that white dwarfs in binary systems, which are thought to be the
progenitors of SNIa supernovae, never have magnetic fields of such
an intensity. Since pre-supernova systems samples are highly
incomplete, it is difficult to address the likelihood of this
conjecture (see below). The second possibility is that asymmetries
do not develop that much because the saturation scale
$\l_{sat} \gg 20\,\delta_{f}$ in these systems. This possibility cannot be
totally disregarded since the saturation scale may be
substantially  different in these systems from that obtained in
laboratory deflagration fronts. The third way out would be that
even if the explosion itself could be largely asymmetric, the
observed remnant could be efficiently symmetrized in the very
early phases of the expansion. Nevertheless, it would be worth to
search for a subset of very asymmetric remnants, perhaps not even
identified as such because a jet-like morphology has never been
expected.

On the other hand, if the surface magnetic field is reduced to
more moderate values ($\lesssim 5 \times 10^7$ G), we have found
that the total asymmetry drops (for saturation scales
$\sim 20 \,\delta_f$) to a value of $\sim 50 \%$ which is much more in
agreement with the asymmetries inferred from present observations.
For instance, for the type Ia supernova 1999by, the first clear
example of an object of that class with asymmetric structure, an
intrinsic polarization of $\sim 0.8 \%$ was observed that was
interpreted as produced by radiation from an oblate or prolate
distribution of scattering electrons with an asymmetry between
$17 \%$ and $50 \%$, depending on the inclination of the object
relative to the line of sight (Howell et al. 2001). Another
important aspect to be noted is that in SN1999by, Si lines
stronger than in normal type Ia SNs have been detected. This could
be explained, in the framework of our model, as a result of an
incomplete burning in the equator direction followed by a smaller
production of Ni which, in turn, could explain the observed redder
light curve. Yet another possibility  may be that the prolate
flame produces a transition to detonation first on the polar
direction, thus disrupting the star before the equatorial
direction had the chance to reach the requirements for detonation
transition, therefore resulting in an incomplete burning and
smaller Ni production.

So far, only $\sim 15 \%$ of the type Ia supernovae  have been
examined through polarimetric observations and they do not
present, in general,  high polarization degrees. The model here
described, could, in principle, explain the asymmetry inferred
from these spectropolarimetric observations, if the fraction of
strongly magnetized white dwarfs (WD) is not too large. In fact,
it seems that only $5 \%$ of the isolated  white dwarfs are
magnetized, but this figure increases to $25 \%$ in binary
systems.

Very recently, Cumming (2002) has suggested that due to an effect
of screening and magnetic field amplification   in rapidly
accreting systems, an asymmetry (due to magnetic field effects)
can be more likely observed in super-soft X-ray binaries. This
constitutes a new (and separate) observational constraint to the
study and detection of the asymmetry effect presented here. In
fact, the search of asymmetric SNIas in super-soft X-ray binaries
could be an easier task  than to survey a full sample  of very
young type Ia supernova remnants through spectropolarimetric
observations.

In the model proposed here, if  higher expansion coefficients,
$a(t)$, were considered, then either a higher magnetic field
strength or a lower saturation scale would be needed in order to
produce the same value for the total asymmetry. In other words, we
are not able to establish completely  a set of parameters for the
model, and a statistically  significant sample of asymmetric type
Ia SNs would be  required for that purpose. It must be also
pointed out that in a real explosion, $a(t)$ should assume a
distinct functional dependence on the pole and the equator, and
this  would enhance the asymmetry effect. In the model presented
here, there is, in fact, a spherical symmetric remnant surrounding
a prolate flame.

We have assumed in the model a constant fractal dimension all over
the explosion. This approximation is  justified by the very fast
growth of the instabilities in the thermonuclear front, so that
the turbulent regime is achieved well  before an appreciable
expansion of the star.

Finally, we should mention that other models have been suggested
in the literature to explain the asymmetries in type Ia supernovae
explosions, although all of them have their problems (see e.g.,
Howell et al 2001). The model here investigated also presents
limitations that could, at least in part, be solved with the help
of future multidimensional numerical calculations and larger
samples of observed asymmetric astrophysical objects. Laboratory
experiments of the deflagration of mixtures of gases in the
presence of magnetic fields could also provide alternative tests
of the asymmetry effect in the flame front and also of the fractal
model examined. The configuration of these experiments must be
settled carefully in order to ensure the  growth of the R-T
instability over a wide spectrum of perturbations, and the
geometry of the apparatus must be chosen in such a way that its
minimum size is larger than the minimum wavelength of the
perturbations (Ghezzi 2002).\footnote{ We  are of course supposing
that the fractal model can
 explain the turbulent regime of a supernova flame.
Previous laboratory experiments
(Gostintesv et al 1988) and numerical
simulations of fractal flames in the wrinkled regime (Filyand et al. 1994; Blinnikov et al. 1996) seem
to support this  hypothesis.}

\section*{Acknowledgements}

We acknowledge partial support of the Brazilian Foundations FAPESP
(S\~ao Paulo) and CNPq to this work.

\bsp
\label{lastpage}

\end{document}